\documentclass[conference]{IEEEtran}
\IEEEoverridecommandlockouts
\usepackage{cite}
\usepackage{amsmath,amssymb,amsfonts}
\usepackage{algorithmic}
\usepackage{graphicx}
\usepackage{textcomp}
\usepackage{xcolor}
\usepackage{url}
\usepackage{orcidlink}

\hypersetup{hidelinks,
	colorlinks=true,
	allcolors=black,
	pdfstartview=Fit,
	breaklinks=true}
\def\BibTeX{{\rm B\kern-.05em{\sc i\kern-.025em b}\kern-.08em
    T\kern-.1667em\lower.7ex\hbox{E}\kern-.125emX}}
\begin{document}
\bibliographystyle{abbrv}

\title{Optimization of Armv9 architecture general large language model inference performance based on Llama.cpp\\
{\footnotesize Take Qwen and Yitian 710 as examples}
}


\author{
	\IEEEauthorblockN{
		Longhao Chen\orcidlink{0009-0008-7803-8208}\IEEEauthorrefmark{1}, 
		Yina Zhao\orcidlink{0009-0003-4591-157X}\IEEEauthorrefmark{2}, 
		Qiangjun Xie\IEEEauthorrefmark{3},
		Qinghua Sheng\IEEEauthorrefmark{3}} 
	\IEEEauthorblockA{\IEEEauthorrefmark{1}CEATRG, Hangzhou Dianzi University, Hangzhou, CN\\ Email: Longhao.Chen@outlook.com}
	\IEEEauthorblockA{\IEEEauthorrefmark{2}CEATRG, Wuhan University, Wuhan, CN\\ Email: 2022302141057@whu.edu.cn}
	\IEEEauthorblockA{\IEEEauthorrefmark{3}Hangzhou Dianzi University, Hangzhou, CN\\ Email: qjunxie@163.com} 
	\IEEEauthorblockA{\IEEEauthorrefmark{3}Hangzhou Dianzi University, Hangzhou, CN\\ Email: sheng7@hdu.edu.cn}
}

\maketitle

\begin{abstract}
This article optimizes the inference performance of the Qwen-1.8B\cite{qwen} model by performing Int8 quantization, vectorizing some operators in llama.cpp\cite{llamacpp}, and modifying the compilation script to improve the compiler optimization level. On the Yitian 710 experimental platform, the prefill performance is increased by 1.6 times, the decoding performance is increased by 24 times, the memory usage is reduced to 1/5 of the original, and the accuracy loss is almost negligible.
\end{abstract}

\begin{IEEEkeywords}
LLM, inference, ARM, llama.cpp, Optimization
\end{IEEEkeywords}

\section{Introduction}
Large Language Models (LLMs) have achieved remarkable performance in most natural language processing downstream tasks, such as text understanding, text generation, machine translation, and interactive Q\&A. However, the billions or even trillions of model parameters pose significant challenges for efficient deployment of LLMs at the edge. With the growth rate of model parameters far outpacing the improvement in hardware performance, the academic and industrial communities are exploring software and hardware collaborative methods like model compression, dataflow optimization and operator invocation to deploy and run large models under the limited hardware conditions. 

This article uses the default quantizer of llama.cpp to perform Int8 quantization on the Qwen-1.8B model, uses ARM's NEON instructions to vectorize some operators in llama.cpp, and modifies the compilation script to improve the GCC compiler optimization level. In the test, the prefill performance increased from 86 token/s to 145 token/s, the decode performance increased from 2 token/s to 48 token/s, the memory usage was reduced from 10GiB to 2.3GiB, and tested using the \verb|piqa|\cite{piqa} data set in \verb|lm-evaluation-harne|\cite{evalharness}, the accuracy is only reduced by 0.0076. All codes in this article are open source to \url{https://github.com/Longhao-Chen/Aicas2024}

\section{Experiment platform}

\subsection{Yitian 710 processor}
Yitian 710 is the first Arm server chip released by T-Head. It is independently designed and developed by T-Head. It adopts an advanced architecture and has the characteristics of high energy efficiency, high bandwidth and is compatible with the Armv9 architecture.

In the Armv9 architecture, VDOT, MMLA and other instructions for Int8 calculations are provided, which can greatly accelerate Int8 type model inference.

\subsection{Qwen}
Qwen is a LLM officially released by Alibaba Cloud Computing Co. Ltd., with parameter scales ranging from billions to trillions. The comprehensive performance of this model is well-rounded in mainstream benchmark evaluations.

In this article, we use Qwen 1.8B with 24 decode layers as an experimental model.

\section{Optimization}
\subsection{Use the latest compiler}
Old compilers do not support Integer Matrix Multiply intrinsics, specifically \verb|__ARM_FEATURE_MATMUL_INT8| is not defined\cite{armsoftwareLanguageExtensions}. To be able to use Integer Matrix Multiply intrinsics we need to use a newer compiler. We are using GCC 13.2.0. In this version, Integer Matrix Multiply intrinsics are fully supported.

\subsection{Use more compiler optimizations}
Compiler optimization is an effective and convenient optimization method, and the GCC compiler supports multiple optimization levels. By default, llama.cpp uses \verb|-O3| optimization. To use higher-level optimization methods, you can use the command \verb|LLAMA_FAST=1 make -j8| to compile. This will enable a higher level of \verb|-Ofast| optimization. Compared with \verb|-O3|, \verb|-Ofast| will enable \verb|-fallow-store-data-races|, \verb|-fassociative-math|, \verb|-fcx-limited-range|, \verb|-fexcess-precision=fast|, \verb|-ffinite-math-only|, \verb|-freciprocal-math|, \verb|-funsafe-math-optimizations|, and disable \verb|-fsemantic-interposition|, \verb|-fsigned-zeros|, \verb|-fmath-errno|, \verb|-ftrapping-math|

Link Time Optimization (LTO) gives GCC the capability of dumping its internal representation (GIMPLE) to disk, so that all the different compilation units that make up a single executable can be optimized as a single module. This expands the scope of inter-procedural optimizations to encompass the whole program (or, rather, everything that is visible at link time)\cite{LTO}. Therefore, for programs composed of multiple source files, enabling LTO can achieve better optimization. In GCC, we can enable LTO by passing \verb|-flto|.

\subsection{Select the architecture of your host system}
Modern cloud computing facilities usually use virtual machines or containers to isolate processes of different users. In these virtual machines or containers, programs often cannot obtain information about the underlying hardware. Therefore, at compile time, the host's architecture needs to be explicitly specified to the compiler. For Yitian 710 processor, you can use \verb|-mcpu=neoverse-n2 -mtune=neoverse-n2| \cite{gnuAArch64Options}.

\subsection{Rewrite some operators using NEON}
Arm Neon technology is an advanced Single Instruction Multiple Data (SIMD) architecture extension for the A-profile and R-profile processors. Neon technology is a packed SIMD architecture. Neon registers are considered as vectors of elements of the same data type, with Neon instructions operating on multiple elements simultaneously. Multiple data types are supported by the technology, including floating-point and integer operations \cite{armNeon}. For simple functions, the compiler's Auto-vectorization can already generate high-performance NEON instructions, but complex functions require manual writing of NEON instructions. We rewrote functions \verb|ggml_fp16_to_fp32_row|, \verb|ggml_fp32_to_fp16_row|, \verb|ggml_compute_forward_norm_f32|, \verb|ggml_compute_forward_rms_norm_f32|, \verb|ggml_compute_fp16_to_fp32|.

\subsection{Reduce unnecessary type conversions}
The 8-bit quantization of llama.cpp treats 32 data as a group by default, and each group uses the same scaling factor. This scaling factor is saved as a \verb|float16| type\cite{githubLlamacppggmlcommonh132f55795e51094954f1b1f647f97648be724a3a}. Limited by processor instruction set, in each calculation, the corresponding scaling coefficient needs to be converted into data types, which will take up part of the time. If you directly use the float32 type to save the scaling factor, no conversion is required, which will save more time. But the memory usage will increase 5.88\% .

\section{Evaluate}
We use the speed of the inference interface in the Python package \verb|transformers 4.38.2| as a baseline.

The \verb|llama.cpp| item in the table is the unmodified original program.

\begin{table}[htbp]
\caption{Precision}
\begin{center}
\begin{tabular}{|c|c|}
\hline
\textbf{Test items} & \textbf{\textit{Accuracy(piqa)}} \\
\hline
Baseline & 0.7312 \\
\hline
llama.cpp + fp16$^{\mathrm{a}}$ & 0.7252 \\
\hline
llama.cpp + Int8$^{\mathrm{a}}$ & 0.7236 \\
\hline
llama.cpp + fp16$^{\mathrm{b}}$ & 0.7252 \\
\hline
llama.cpp + Int8$^{\mathrm{b}}$ & 0.7236 \\
\hline
Ours & 0.7236 \\
\hline
\multicolumn{2}{l}{$^{\mathrm{a}}$Compiler: gcc 9.4.0}\\
\multicolumn{2}{l}{$^{\mathrm{b}}$Compiler: gcc 13.2.0}
\end{tabular}
\label{tab1}
\end{center}
\end{table}

\begin{table}[htbp]
\caption{Memory usage}
\begin{center}
\begin{tabular}{|c|c|c|}
\hline
\textbf{Test items} & \textbf{\textit{Physical memory (MiB)}}& \textbf{\textit{Virtual memory (MiB)}} \\
\hline
Baseline & 10627 & 12228 \\
\hline
llama.cpp + fp16$^{\mathrm{a}}$ & 3807 & 4204 \\
\hline
llama.cpp + Int8$^{\mathrm{a}}$ & 2165 & 2562 \\
\hline
llama.cpp + fp16$^{\mathrm{b}}$ & 3807 & 4205 \\
\hline
llama.cpp + Int8$^{\mathrm{b}}$ & 2166 & 2563 \\
\hline
Ours &  2250 & 2649 \\
\hline
\multicolumn{3}{l}{$^{\mathrm{a}}$Compiler: gcc 9.4.0}\\
\multicolumn{3}{l}{$^{\mathrm{b}}$Compiler: gcc 13.2.0}
\end{tabular}
\label{tab1}
\end{center}
\end{table}

\begin{table}[htbp]
\caption{Inference rate}
\begin{center}
\begin{tabular}{|c|c|c|}
\hline
\textbf{Test items} & \textbf{\textit{Prefill rate (tokens/s)}} & \textbf{\textit{Decode rate (tokens/s)}}\\
\hline
Baseline &  86.79 & 2.07 \\
\hline
llama.cpp + fp16$^{\mathrm{a}}$ & 113.63 & 24.05 \\
\hline
llama.cpp + Int8$^{\mathrm{a}}$ & 38.53 & 24.36 \\
\hline
llama.cpp + fp16$^{\mathrm{b}}$ & 116.85 & 23.51 \\
\hline
llama.cpp + Int8$^{\mathrm{b}}$ & 98.55 & 37.88 \\
\hline
Ours & 145.86 & 48.36 \\
\hline
\multicolumn{3}{l}{$^{\mathrm{a}}$Compiler: gcc 9.4.0}\\
\multicolumn{3}{l}{$^{\mathrm{b}}$Compiler: gcc 13.2.0}
\end{tabular}
\label{tab1}
\end{center}
\end{table}
\section{Discussion}
Through experimental data, it can be seen that our solution greatly improves the inference performance of large language models with a low accuracy drop. Next, we can consider using the float type to save the scaling factor during quantization, which may result in smaller accuracy loss.

\section*{Acknowledgment}
I would like to thank my parents and school for their support, and I would also like to thank the conference organizers for their support.

\bibliography{bibfile.bib}

\end{document}